\documentclass[twocolumn,showpacs,showkeys,preprintnumbers,amsmath,amssymb]{revtex4}

\usepackage{graphicx}
\usepackage{dcolumn}
\usepackage{bm}

\newcommand{\bv}[1]{\mbox{\boldmath$#1$}}

\begin{document}

\preprint{ }

\title{A novel method to create a vortex in a Bose-Einstein condensate}

\author{Shin-Ichiro Ogawa, Mikko M\"ott\"onen$^1$,
Mikio Nakahara$^{1,2}$, Tetsuo Ohmi$^{3}$,
and Hisanori Shimada$^{2}$\\}

\affiliation{%
Department of Physics, Osaka City University, Osaka 558-8585, Japan\\
$^1$Materials Physics Laboratory, Helsinki University of Technology, P.O. Box 2200 (Technical Physics), FIN-02015 HUT Finland\\
$^2$Department of Physics, Kinki University, Higashi-Osaka 577-8502, Japan\\
$^3$Department of Physics, Kyoto University, Kyoto 606-8502, Japan
}%

\date{\today}

\begin{abstract}
It has been shown that a vortex in a BEC with spin degrees of freedom
can be created by manipulating with external magnetic
fields. In the previous work, an optical plug along the vortex axis 
has been introduced to avoid Majorana flips, which take place when the 
external magnetic field vanishes along
the vortex axis while it is created. In the present work, in contrast, we
study the same scenario without introducing the optical plug.
The magnetic field vanishes only in the center of the vortex
at a certain moment of the evolution and hence we expect that the system will
lose only a fraction of the atoms by Majorana flips even
in the absence of an optical plug.
Our conjecture is justified by numerically solving the
Gross-Pitaevskii equation, where the full spinor degrees of freedom of the
order parameter are properly taken into account.
A significant simplification of the experimental realization of the scenario
is attained by the omission of the optical plug.
\end{abstract}

\maketitle

\section{Introduction}

Alkali atoms become a superfluid upon
Bose-Einstein condensation (BEC) [1, 2].
The superfluid properties of the system were considered to be
essentially the same as those of superfluid $^4$He, in spite
of the fact that the former is a weakly-coupled system while the latter
is coupled strongly. In contrast with $^4$He, however, alkali atoms
have internal degrees of freedom attributed to the
hyperfine spin ${\bv{F}}$ and, accordingly,
the order parameter has $2|F|+1$ components [3, 4]. The atoms
$^{23}$Na and $^{87}$Rb have $|F|=1$, for example, and their order parameters
have three components, similarly to the orbital or the spin part of
the superfluid $^3$He.

These degrees of freedom bring about remarkable
differences between the BEC of alkali atoms and that of $^4$He.
The hyperfine spin freezes along the direction of the local magnetic
field when a BEC is magnetically trapped.
If a BEC is trapped optically, on the other hand, these degrees of
freedom manifest themselves and various new phenomena, such as the phase
separation between the different spin states that have never been seen
in superfluid $^4$He, may be observed.

It was suggested in [5] and [6] that a totally new process for 
formation of a vortex in a BEC of alkali atoms
is possible by making use of the hyperfine
degrees of freedom to ``control'' the BEC.
Suppose a BEC is confined in a Ioffe-Pritchard trap.
Then a vortex state with two units of circulation can be continuously
created from the vortex free superfluid state by simply reversing
the axial magnetic field $B_z$, parallel to the Ioffe bars.
This field is created by a set of pinch coils and can be easily controlled.
Four groups have already reported the formation of vortices
with three independent methods [7-11].
Our method is totally different from the previous ones in
that the hyperfine spin degrees of freedom have been fully utilized.
The present paper is a sequel to [6]. Here, we conduct
further investigations
on the formation of a vortex by reversing the axial
field $B_z$. An optical plug along the axis was introduced in [5] and [6]
to avoid possible Majorana flips
near the axis, that may take place when $B_z$ passes through zero.
It is expected, however, that trap loss due to Majorana flips
may not be very significant if the
``dangerous point'' $|B|=0$ is passed fast enough
and that considerable amount of the BEC remains in the trap as the result.
This process should not
be too fast, however, such that the adiabatic condition is still satisfied.
It is a very difficult task to introduce a sharply forcused
optical plug along the center of the condensate whose radius
is of the order of a few microns. Accordingly we expect that experimental 
realization of our scenario will be much easier without the
optical plug. In the present paper, therefore, we analyze our scenario by
numerically integrating the multi-component Gross-Pitaevskii equation.

This paper is organized as follows. In Sec. II, we outline
the order parameter and
the Gross-Pitaevskii equation for an $|F|=1$ BEC.
In Sec. III, the Gross-Pitaevskii equation is integrated numerically
to analyze the time dependence of the order parameter. It will be shown
that merely half of the condensate is lost from the trap if the
time dependence of the external magnetic field is chosen properly.
Sec. IV is devoted to conclusions and discussions.
In the Appendix, it is shown that the formation of a vortex
in the present scenario may be understood in terms of the Berry phase.

\section{Order Parameter and Gross-Pitaevskii Equation of $|F|=1$ BEC}

Let us briefly summarize the order parameter and the
Gross-Pitaevskii equation for a BEC with the
hyperfine spin $F=1$ to make this paper self-contained. 
The readers should be referred to [3] and [6] for further details. 

The order parameter of a BEC with $|F|=1$
has three components $\Psi_i\ (i=-1, 0, +1)$ with respect to the
basis vectors $| i \rangle$ defined by $F_z| i\rangle = i|i\rangle$.
The order parameter $|\Psi \rangle$ is then expanded as
$$
|\Psi \rangle = \sum_{i=0, \pm 1} \Psi_i| i \rangle.
$$
It turns out, however, that another set of basis vectors $|a 
\rangle\ (a=x,y,z)$, defined by $F_{a}|a\rangle=0$, is
more convenient for certain purposes. The order parameter is now expressed
as
$$
|\Psi \rangle =\sum_{a = x,y,z} \Psi_{a} |a \rangle.
$$
The transformation matrix from $\{\Psi_i\}$ to $\{\Psi_{a}\}$
is found in [6]. 

The most general form of the Hamiltonian for $|F|=1$ atoms,
that is rotationally invariant except for the Zeeman term, is
\begin{eqnarray}
\hat{H}-\mu \hat{N} &=& \int \Big[ 
\psi^{\dagger}_{a}\left(-\frac{\hbar^2}{2m}\nabla^2-\mu \right)
\psi_{a} + \frac{g_1}{2} (\psi^{\dagger}_{a} \psi_{a})^2
\nonumber\\
& &+ \frac{g_2}{2} |\psi_{a} \psi_{a}|^2 
+ i \varepsilon_{a b c}\omega_{L c}
\psi^{\dagger}_{a} \psi_{b}\Big] d^3\bv{r},
\end{eqnarray}
where $\hbar \omega_{L a} = \gamma_{\mu} B_{a}$, $\gamma_{\mu}
\simeq \mu_B/2$
being the gyromagnetic ratio of the atom. The coupling constants
are given by
\begin{equation}
g_1 = \frac{4 \pi \hbar^2}{m} a_{2},\quad g_2 = \frac{4 \pi \hbar^2}{m} 
\frac{a_0-a_2}{3},
\end{equation}
where $a_2 = 2.75\;\mathrm{nm}$ and $a_0 = 2.46\;\mathrm{nm}$ [12].

The Heisenberg equation of motion derived from the above Hamiltonian is
\begin{eqnarray}
i \hbar 
\frac{\partial \psi_{a}}{\partial t} &=&[\psi_{a}, \hat{H}-\mu \hat{N}]
\nonumber\\
&=& \left(-\frac{\hbar^2}{2m} \nabla^2-\mu \right) \psi_{a} + g_1
(\psi^{\dagger}_{b} \psi_{b}) \psi_{a}\nonumber\\
& &+ g_2 (\psi_{b} \psi_{b}) \psi_{a}^{\dagger} + 
i \varepsilon_{a b c} \omega_{L c} \psi_{b}.
\end{eqnarray}
By taking the expectation value of the above equation in the 
mean-field approximation, $\langle \psi \psi \psi \rangle \simeq \Psi \Psi
\Psi$, where $\Psi_{a} = \langle \psi_{a} \rangle$, we obtain
\begin{eqnarray}
i \hbar \frac{\partial \Psi_{a}}{\partial t} &=& 
\left( -\frac{\hbar^2}{2m} \nabla^2-\mu \right) \Psi_{a}
+ g_1|\Psi|^2 \Psi_{a}
\nonumber\\
& & + g_2 (\Psi)^2 \Psi^*_{a}+ i \varepsilon_{abc}
\omega_{Lc} \Psi_{b}.
\label{gp}
\end{eqnarray}
This is the fundamental equation which we will use in the rest of this paper.

\section{Creation of a Vortex}

\subsection{Magnetic Fields}

Suppose we confine a BEC in a Ioffe-Pritchard trap
with a quadrupole magnetic field
\begin{equation}
\bv{B}_{\perp}(\bv{r}) = \left(
\begin{array}{c}
B_{\perp}(r) \cos(-\phi)\\
B_{\perp}(r) \sin (-\phi)\\
0
\end{array} \right)
\end{equation}
and a uniform axial field 
\begin{equation}
\bv{B}_{z}(t) = \left(
\begin{array}{c}
0\\
0\\
B_z(t)
\end{array} \right).
\end{equation}
Here $(r, \phi, z)$ are the cylindrical coordinates.
The magnitude $B_{\perp}(r)$ is proportional to $r$ near the axis $r \sim 0$;
$B_{\perp}(r) \simeq B' r$, with $B'$ being a constant.
The system is assumed to be uniform along the $z$ direction for
calculational simplicity. Our analysis should apply to a cigar-shaped
system as well.

Suppose we reverse the $\bv{B}_z$ field slowly,
while keeping $\bv{B}_{\perp}$ fixed as shown in Fig. 1. 
It was shown in [5, 6] that a vortex with two units of circulation
will be formed if we start with a vortex-free BEC.
In these papers,
an optical plug was introduced along the vortex axis
to prevent the atoms from escaping from the trap when
$\bv{B} = \bv{B}_{\perp} + \bv{B}_z$ vanishes at $r=0, t= T/2$.
Accordingly, the order parameter remains within the weak-field
seeking state (WFSS) throughout the scenario.
\\
\begin{figure}
\begin{center}
\includegraphics[width=5cm]{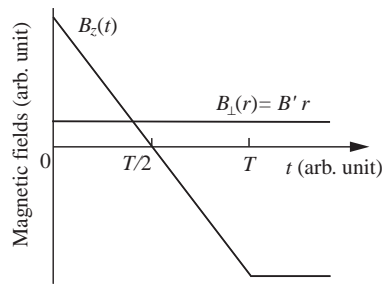}
\end{center}
\caption{The schematic time dependence of the axial field $B_z$.
The field $B_z$ reverses the sign in an interval of time $T$,
while the quadrupole field $\bv{B}_{\perp}=B' r$ remains fixed. The field
$B_z(t)$ is fixed at the value $B_z(T)=-B_z(0)$ for $t>T$ for
further evolution.}
\label{fig:1}
\end{figure}

We suspect, however, that the BEC may be stable even without
the optical plug since $\bv{B}$ vanishes only in $r=0$ at the
time $t=T/2$. This will be justified by solving the Gross-Pitaevskii
equation numerically below.
In contrast with the previous work,
we have to take the full degrees of freedom of the order
parameter into account since the energies of the
three hyperfine states (the strong-field
seeking state, the weak-field seeking state and the neutral state)
are degenerate when $\bv{B}=0$.

\subsection{Initial State}

We first solve the Gross-Pitaevskii equation in the stationary state to find
the initial order parameter configuration and the corresponding
chemical potential. There is a sufficient gap
between the weak-field seeking state
and the other states at $t=0$ and the BEC
may be assumed to be purely in the weak-field seeking state.
Let us parametrize the external magnetic field $\bv{B}$ with use of the polar 
angles $\alpha$ and $\beta$ as
\begin{equation}
\bv{B}(r, t) = \bv{B}_{\perp}(r) + \bv{B}_z(t) =|\bv{B}|
\left( \begin{array}{c}
\sin \beta \cos \alpha\\
\sin \beta \sin \alpha\\
\cos \beta
\end{array} \right),
\end{equation}
where $\alpha = -\phi$ for the quadrupole field and
\begin{equation}
\beta = \tan^{-1} \left[\frac{|\bv{B}_{\perp}(r)|}{|\bv{B}_z(t) |}\right].
\end{equation}
Since the hyperfine spin $\bv{F}$ is 
antiparallel with $\bv{B}$, we must have
\begin{equation}
\hat{\bv{l}} =
\left( \begin{array}{c}
-\sin \beta \cos \alpha\\
-\sin \beta \sin \alpha\\
-\cos \beta
\end{array} \right),
\end{equation}
where $\hat{\bv{l}}$ is a unit vector parallel to $\bv{F}$, see Fig. 2.

\begin{figure}
\begin{center}
\includegraphics[width=5cm]{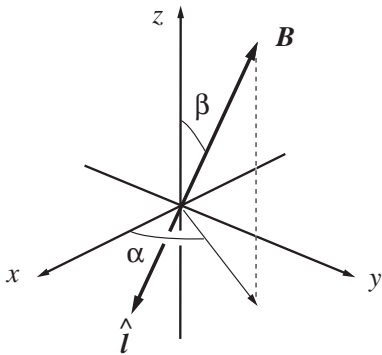}
\end{center}
\caption{Polar angles $(\alpha, \beta)$
parametrizing $\bv{B}$ and $\hat{\bv{l}} \parallel \bv{F}$.}
\label{fig:2}
\end{figure}

The most general form for
the order parameter yielding the above $\hat{\bv{l}}$ vector is
\begin{equation}
\Psi_{a} = \frac{f_0}{\sqrt{2}} e^{-i \gamma}\left( \begin{array}{c}
\cos \beta \cos \alpha + i \sin \alpha\\
\cos \beta \sin \alpha - i \cos \alpha\\
-\sin \beta
\end{array} \right) \equiv f_0 v_{a},
\label{wfss}
\end{equation}
where $\{v_{a}\}$ represents the ``phase'',
while $f_0$ is the amplitude of the order parameter.
In mathematical terms, $\{v_{a}\}$ defines the local $SO(3)$ frame or the
``triad'' of the real orthonormal vectors $\{\hat{\bv{m}}, \hat{\bv{n}}, 
\hat{\bv{l}\}}$, where
\begin{equation}
v_{a} = \frac{1}{\sqrt{2}}e^{-i\gamma} (\hat{\bv{m}}+i \hat{\bv{n}})_{a}, \quad \hat{\bv{l}} = \hat{\bv{m}} \times \hat{\bv{n}}.
\end{equation}
It should be noted that the above order parameter takes the same form as
that of the orbital part of the superfluid $^3$He-A.

One may obtain more insight if the order parameter in Eq.~(\ref{wfss}) is rewritten
in the $\Psi_i$ basis as [13]
\begin{equation}
\Psi_{i} =f_0 e^{-i \gamma}
\left( \begin{array}{c}
\frac{1}{2} e^{-i \alpha}(1-\cos \beta)\\
-\frac{1}{\sqrt{2}} \sin \beta\\
\frac{1}{2} e^{i\alpha}(1+\cos \beta)
\end{array}
\right).
\end{equation} 
The angle $\beta$ vanishes at $r=0$ (note that $\bv{B}_{\perp}=0$ at $r=0$),
hence
we find $\Psi_{+1}=
\Psi_{0} = 0$ while $\Psi_{-1} \propto e^{i(\alpha-\gamma)} \neq 0$ there. 
For the order parameter to be smooth at $r=0$, we have to choose
\begin{equation}
\gamma = \alpha = -\phi.
\end{equation}
Accordingly, the order parameter takes the form
\begin{equation}
\Psi_{a} = \frac{f_0}{\sqrt{2}} e^{i \phi}\left( \begin{array}{c}
\cos \beta \cos \phi - i \sin \phi\\
-\cos \beta \sin \phi - i \cos \phi\\
-\sin \beta
\end{array} \right)
\label{wfa}
\end{equation}
in the $\Psi_{a}$ basis and
\begin{equation}
\Psi_{i}=\frac{f_0}{2}\left( \begin{array}{c}
(1-\cos \beta)e^{2i \phi}\\
-{\sqrt{2}} \sin \beta e^{i \phi}\\
1+\cos \beta
\end{array} \right)
\label{wfi}
\end{equation}
in the $\Psi_i$ basis. It follows from Eq.~(\ref{wfi}) that the state with
$\hat{\bv{l}}=-\hat{\bv{z}}\ (\beta=0)$ has a vanishing winding number,
while that with $\hat{\bv{l}}=\hat{\bv{z}}\ (\beta=\pi)$ has the winding
number $2$. Accordingly, the former state may be continuously deformed
to the latter state by changing $\beta$ from $0$ to $\pi$ smoothly,
resulting in the formation of a vortex
with winding number 2 [14, 15]. It is shown in the
Appendix that this phase may also
be understood in terms of the Berry phase associated with the
adiabatic change of the local magnetic field.

The stationary Gross-Pitaevskii equation is given by
\begin{equation}
-\frac{\nabla^2}{2m} \Psi_{a} + g_1 |\Psi|^2 \Psi_{a} + g_2 \Psi^2 \Psi_a^*
+ i  \varepsilon_{a b c} \Psi_{b}\omega_{L c}
= \mu \Psi_{a}.
\label{gps}
\end{equation}
By substituting the WFSS order parameter in Eq.~(\ref{wfa}) into Eq.~(\ref{gps}), 
we obtain
\begin{eqnarray}
&&-\frac{\hbar^2}{2m}\Big[\nabla^2 -\frac{\beta^{\prime 2}}{2}-
\frac{1}{4r^2}(7-8 \cos \beta + \cos 2\beta)\Big] f_0\qquad \nonumber\\
& &\qquad \qquad + \gamma_{\mu} B(r) f_0 + g_1 f_0^3 = \mu f_0.
\end{eqnarray}
Note that there appears an extra term $\beta^{\prime 2}$ that is missing if
this BEC is described by a scalar (namely $U(1)$) order parameter. This term
originates from $v_{a}^{\dagger} \nabla^2 v_{a}$ and
represents the rotation of the local $SO(3)$ frame. In ordinary physical
settings, however, this term is smaller than the $B$-term by a factor of
$\omega/\omega_L \sim 10^{-3}$ and hence its effect may be negligible
in the following arguments.

The amplitude of the external magnetic field has an
approximately harmonic
potential profile near the origin,
\begin{eqnarray}
B(r,t) &=& 
\sqrt{B_{\perp}^2(r) + B_z^2(t)} \simeq \sqrt{B^{\prime 2} r^2 + B_z^2(t)}\nonumber\\
&\simeq& B_z(t) +\frac{B^{\prime 2}}{2B_z(t)}r^2. 
\end{eqnarray}
It turns out to be convenient to scale the energies by the 
energy-level spacing $\hbar \omega$ at $t=0$ and the
lengths by the harmonic-oscillator length $a_{\rm HO}$, where
\begin{eqnarray}
\omega &=& B' \sqrt{\frac{\gamma_{\mu}}{m B_z(0)}},\\
a_{\rm HO} &=& \sqrt{\frac{\hbar}{m \omega}}.
\end{eqnarray}
If we take $^{23}$Na and substitute $B_z (0)= 1 \;\mathrm{G}$ and $B' = 300
\;\mathrm{G/cm}$, we find $a_{\rm HO} \sim 9.14 \times 10^{-1}\;
\mathrm{\mu m}$
and $\hbar \omega \sim 3.49 \times 10^{-24}\;\mathrm{erg}$.
For the same choice of the parameters, we have $\omega_L(r=0, t=0) \sim 
4.40 \times 10^6\;{\mathrm{rad/s}} \sim 1.330 \times 10^3 \omega$.
It is reasonable to
assume
$\tau \equiv 2\pi/\omega_L(r=0, t=0)
\sim 1.43\;\mathrm{\mu s}$ to be the measure of the adiabaticity.

After these scalings, the dimensionless Gross-Pitaevskii equation takes
a simpler form
\begin{eqnarray}
-\frac{1}{2}\tilde{\nabla}^2 \tilde{f}_0
&+&\left[\tilde{B}(r) +\frac{\beta^{\prime 2}}{4}
 +  \frac{1}{8\tilde{r}^2}
(7-8 \cos \beta + \cos 2\beta)\right] \tilde{f}_0\nonumber\\
& &+ \tilde{g}_1 \tilde{f}_0^3 = \tilde{\mu} \tilde{f}_0,
\end{eqnarray}
where $\tilde{r} =r/a_{\rm HO}, \tilde{f}_0 = f_0 a_{\rm HO}^{3/2}, 
\tilde{\mu} = \mu/\hbar \omega$, $\tilde{B} = \gamma_{\mu}B/\hbar \omega$
and $\tilde{g}_1 = g_1/a_{\rm HO}^3 \hbar \omega \sim 0.0378$.
A similar scaling for $g_2$ yields $\tilde{g}_2 \sim -0.00132$.
Note that the singularity at $\tilde{r}=0$ vanishes if $\beta(\tilde{r})$ 
approaches to 0 fast enough as $\tilde{r} \to 0$. 
The tildes will be dropped hereafter whenever it does not cause
confusion. The eigenvalue $\mu$ may be obtained numerically. For
$^{23}$Na with $B_z$ and $B'$ given above, we find
${\mu} - \hbar \omega_L(r=0, t=0) = 3.66$, which amounts
to $1.28 \times 10^{-23}\;{\mathrm erg}$ in dimensional units.
Figure 3 shows the corresponding condensate profile $f_0(r)$.
\\
\begin{figure}
\begin{center}
\includegraphics[width=6cm]{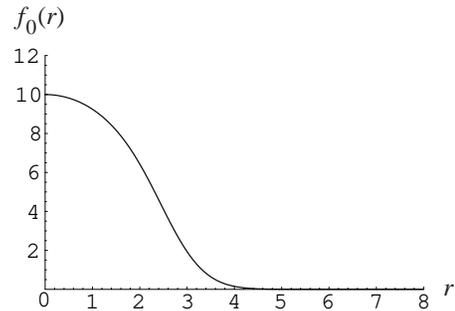}
\end{center}
\caption{Condensate wave function $f_0(r)$.
The condensate wave function and the radial coordinate
are scaled by $a_{\rm HO}^{-3/2}$ and $a_{\rm HO}$, respectively,
and hence dimensionless.}
\label{fig:3}
\end{figure}

\subsection{Time Development}

The time development of the condensate wavefunction is obtained
by solving the Gross-Pitaevskii Eq. (\ref{gp}) which
is written in dimensionless form as
\begin{eqnarray}
i \frac{\partial \Psi_{a}}{\partial t} &=& -\frac{1}{2} \nabla^2 
\Psi_{a}
+ g_1|\Psi|^2 \Psi_{a} \nonumber\\
& &+ g_2 (\Psi)^2 \Psi^*_{a}
+ i \varepsilon_{a b c} \Psi_{b} \omega_{Lc}.
\end{eqnarray}
The initial condition is
\begin{equation}
\Psi_{a} = f_0(r) v_{a},
\end{equation}
where $f_0(r)$ is obtained in the previous subsection
and $v_{a}$ is defined in Eq. (\ref{wfss}).
The condensate cannot remain within the weak-field seeking state during
the formation of a vortex and we have to utilize the full degrees of freedom
of the order parameter $\Psi_{a}$. The Gross-Pitaevskii equation
has been solved numerically to find the temporal development of the order
parameter
while $B_z$ 
changes as
\begin{equation}
B_z(t) = B_z(0) \left[ 1-\frac{2t}{T}\right] \quad (0 \leq t \leq T),
\end{equation}
with several choices for the reversing time $T$.
Adiabaticity is not guaranteed at $t\sim T/2$ when
the energy gaps among the weak-field seeking state (WFSS), the
neutral state (NS) and the strong-field seeking state (SFSS) disappear. 
We expect, however, that this breakdown of 
adiabaticity is not very significant to the condensate
since it takes place only along
the condensate axis for a short period of time.

\begin{figure}
\begin{center}
\includegraphics[width=8.5cm]{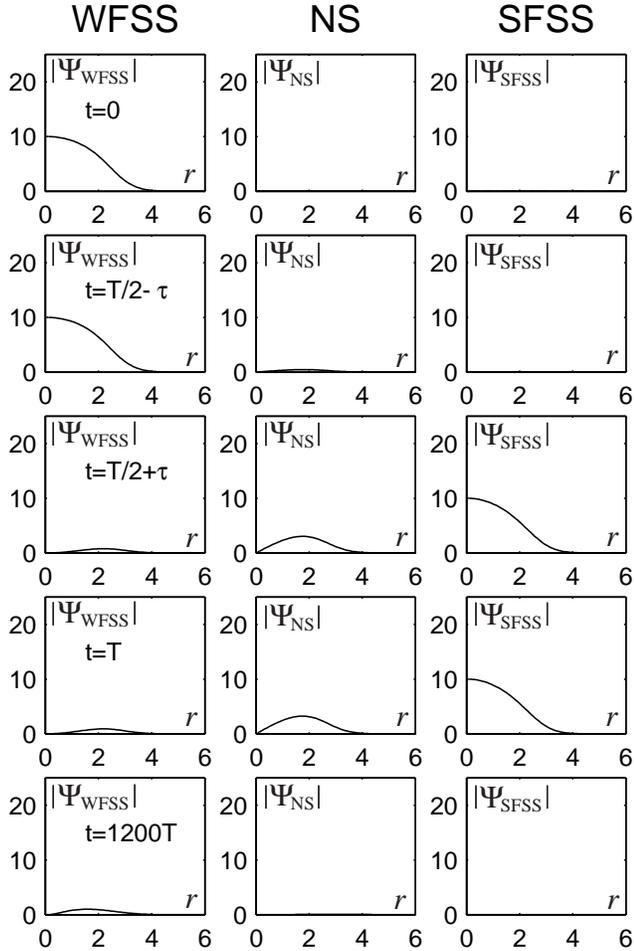}
\end{center}
\caption{Temporal development of the order parameters
$|\Psi_{\rm WFSS}|$, $|\Psi_{\rm NS}|$ and $|\Psi_{\rm SFSS}|$ for
$T= 10 \tau$, where $\tau \sim 1.43 \mu{\rm s}$.
The graphs are plotted for $t=0$ (initial state),
$t=T/2-\tau$ (slightly before $B_z=0$ is crossed), $t=T/2 + \tau$
(slightly after $B_z=0$ is crossed) and $t=T$ ($B_z$ is completely reversed).
Furthermore, $B_z$ is fixed to $B_z(T)=-B_z(0)$ for $t > T$.
The bottom row shows the order parameters at $t=1200 T$.
The order parameters and the radial coordinate are plotted in units of
$a_{\rm HO}^{-3/2}$ and $a_{\rm HO}$, respectively.}
\label{fig:4}
\end{figure}
\begin{figure}
\begin{center}
\includegraphics[width=7cm]{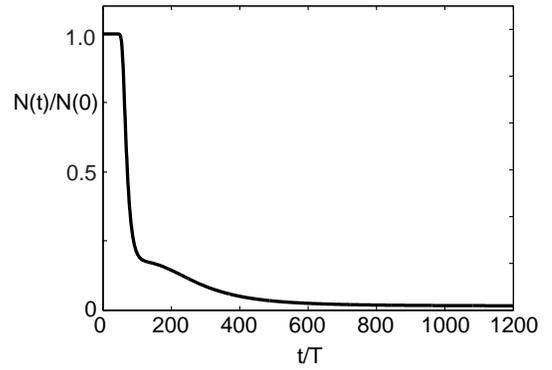}
\end{center}
\caption{The ratio $N(t)/N(0)$
as a function of $t/T$ for the reversing time $T=10 \tau$. Here
$N(t)$ is the number of atoms$/$unit
length along the vortex axis at time $t$. The magnetic 
field $B_z$ decreases monotonically
for $0 \leq t/T \leq 1$, but is kept fixed to $-B_z(0)$ for
$T \leq t$, see Fig. 1.}
\label{fig:5}
\end{figure}
%
\begin{figure}
\begin{center}
\includegraphics[width=8.5cm]{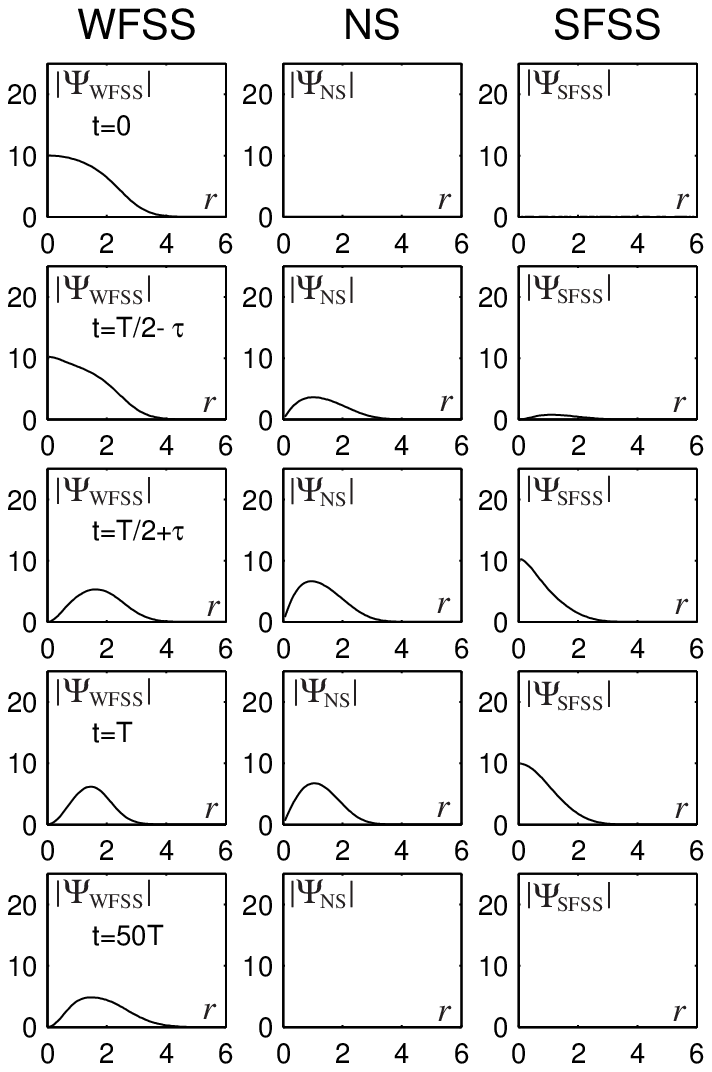}
\end{center}
\caption{Same as in Fig. 4, but for
$T= 100 \tau$, $\tau \sim 1.43 \mu{\rm s}$.
The graphs are plotted for $t=0, t=T/2-\tau,
t=T/2 + \tau, t=T$, and $t=50 T$.}
\label{fig:6}
\end{figure}
\begin{figure}
\begin{center}
\includegraphics[width=7cm]{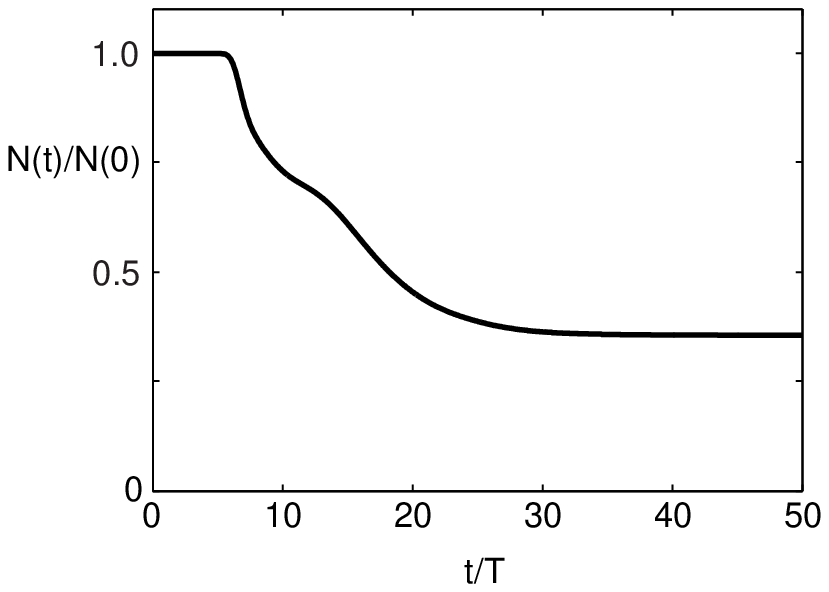}
\end{center}
\caption{Same as in Fig. 5, but for $T=100 \tau$.}
\label{fig:7}
\end{figure}
%
\begin{figure}
\begin{center}
\includegraphics[width=8.5cm]{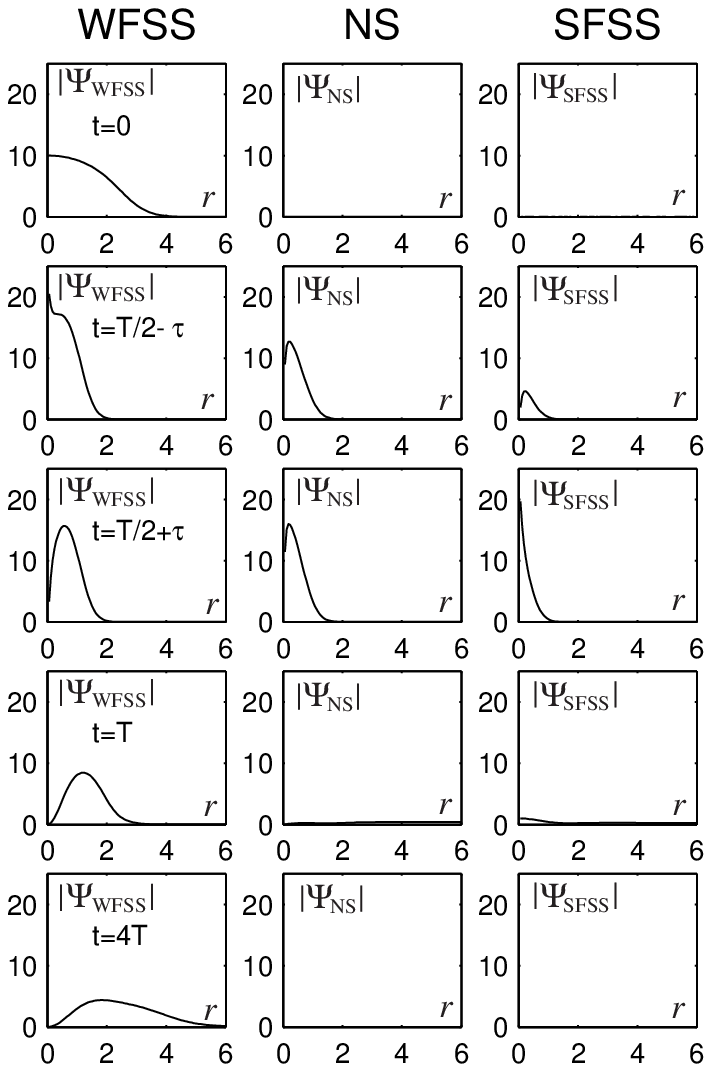}
\end{center}
\caption{Same as in Fig. 4, but for $T= 1000 \tau$, $\tau=1.43\mu{\rm s}$.
The graphs are plotted for $t=0, t=T/2-\tau,
t=T/2 + \tau, t=T$, and $t=4 T$.}
\label{fig:8}
\end{figure}
\begin{figure}
\begin{center}
\includegraphics[width=7cm]{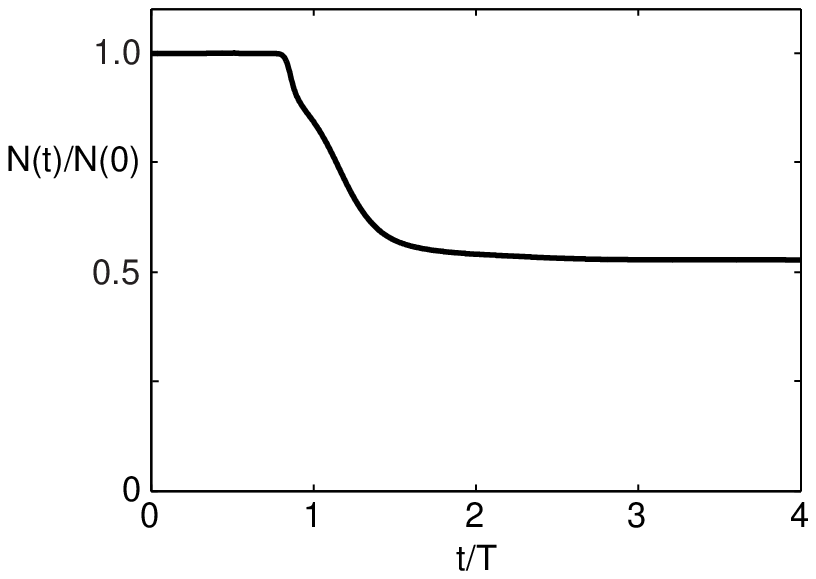}
\end{center}
\caption{Same as in Fig. 5, but for $T=1000\tau$.}
\label{fig:9}
\end{figure}

We project $\Psi_{a}$ thus obtained to the local 
WFSS, NS and SFSS by defining the
projection operators to the respective states as
\begin{equation}
\begin{array}{c}
\Pi_{\rm{W}a b} = v_{a} v_{b}^{\dagger},\quad
\Pi_{\rm{S}ab} = u_{a} u_{b}^{\dagger},
\vspace{0.3cm}\\
\Pi_{\rm{N}ab} = \delta_{ab} - \Pi_{\rm{W}ab} -
 \Pi_{\rm{S}ab}
\end{array}
\end{equation}
where
\begin{equation}
v_{a}= \frac{1}{\sqrt{2}}(\hat{\bv{m}}+i \hat{\bv{n}})_{a}, 
u_{a} = \frac{1}{\sqrt{2}}(\hat{\bv{m}}-i\hat{\bv{n}})_{a}
\end{equation}
define the local WFSS and SFSS $SO(3)$ frames, respectively.

Figures 4, 6, and 8 depict the
amplitudes of the wave functions at $t=0, T/2-\tau, T/2 + \tau$ and $T$
for $T=10 \tau, 100 \tau$, and $1000 \tau$.
Figures 5, 7, and 9 show the particle numbers within the trap for
the same choices of $T$.
To simulate the trap loss, we introduced a function
\begin{equation}
h(r) = \frac{1}{2} \left[ 1 - \tanh\left(\frac{r-r_0}{\lambda} \right)
\right],
\end{equation}
with $r_0 = 30$ and $\lambda=2$. The wave function
$\Psi_{a}(r)$ is multiplied by $h(r)$ after each step
of the Crank-Nicholson algorithm. The parameter $r_0$ roughly corresponds to
the trap size, while $\lambda$ is taken large enough to prevent the
wavefunction being reflected at $r \sim r_0$.
Thus particles that reach beyond $r \sim r_0$ disappear from the trap.
The parameters $r_0$ and $\lambda$ are introduced for purely computational
purposes and should not be confused with any realistic experimental settings.
As $t \gg T$, the condensates in SFSS and NS disappear from the trap and the
particle number reaches its equilibrium value.
The bottom row of Figs. 4, 6, and 8 shows the wavefunctions after
the equilibrium is attained. 

Figure 4 shows the behavior of the condensate when $T=10 \tau$.
In this case,
$B_z(t)$ is reversed so fast that most of the particles
remain in the WFSS at $t =T/2-\tau$ and they are
suddenly converted into the SFSS at
$t=T/2+\tau$. These particles in the SFSS and the NS are eventually 
lost from the trap as $t \to \infty$, see Fig. 5.

Figures 6 and 8 show that the behavior of the condensate is qualitatively
similar for $T=100 \tau$ and $T=1000 \tau$. The condensate is transferred
from the WFSS to the SFSS more efficiently at $t=T/2 -\tau$ for $T=1000 \tau$
which leads to more WFSS components at $t=T/2 +\tau$ in this case.
At a later time $t > T$, 
the condensate is found to oscillate in the trap with the
frequency $\sim \omega$. 

We have also analyzed the case $T=10000 \tau$ and found no qualitative
difference compared to the case $T=1000 \tau$. 
There are slightly more of the NS component in the former case 
at $t=T/2 -\tau$ than the latter case, which leads
to less particles in its equilibrium state at $t\gg T$ . 

It is certainly desirable to have more particles remaining in the trap
when a vortex is created. Figure 10 shows the ratio of the final
particle number to the initial particle number as a function of
the reversing time $T$. Note that the final particle number is evaluated
when the equilibrium is reached.
It is found that a
considerable amount of the condensate $(\gtrsim 1/3)$
is left in the trap for a wide range of the reversing times
$10^2 \lesssim  T/\tau \lesssim 10^4$.
\begin{figure}
\begin{center}
\includegraphics[width=8cm]{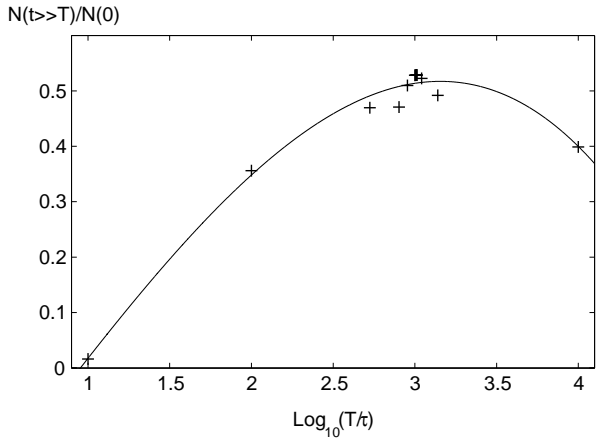}
\end{center}
\caption{Ratio of the residual particle number to the initial 
particle number as a function of $\log_{10}(T/\tau)$.
The crosses are the results of our numerical calculation while
the interpolating curve is introduced as a guide.}
\label{fig:10}
\end{figure}

In summary, our numerical calculations indicate that 
a large fraction of the BEC remains in the trap when $B_z$ is
reversed with proper choices of the reversing time $T$. 
Although there remain SFSS and NS components at $t=T$,
they eventually disappear from the trap at $t \gg T$ when the equilibrium
is reached. The condensate is converted into a state with two units of
circulation. Thus we conclude that a
vortex can be created, even in the absence of an optical plug,
by simply reversing the axial field $B_z$.
We expect that this makes the experimental 
realization of the present scenario much easier than that with an
optical plug [5, 6].

\section{Conclusions and Discussions}

We have analyzed a simple scenario of vortex formation in a spinor BEC.
An axial magnetic field is slowly reversed while the quadrupole field
has been kept fixed in a Ioffe-Pritchard trap, which results in a
formation of a vortex with the winding number $2$. The spinor
degrees of freedom have been fully utilized which renders our
scheme much simpler compared with other methods.

Since the vortex thus created has a higher winding number, it is
metastable and eventually breaks up to two singly quantized vortices.
The lifetime of the metastable state is quite an interesting quantity
to evaluate since its magnitude, compared with the field-reversal time $T$
and the trap time, will be crucial to determine the ultimate
fate of the vortex.

A similar analysis with an $F=2$ BEC is under progress and will be
published elsewhere. 

\begin{acknowledgments}
We would like to thank Kazushiga Machida and Tomoya Isoshima
for useful discussions.
One of the authors (MN) would like to thank partial support by
Grant-in-Aid from Ministry of Education, Culture, Sports, Science and
Technology, Japan (Project Nos. 11640361 and 13135215).
He also thanks Martti M. Salomaa for support and
warm hospitality in the
Materials Physics Laboratory at Helsinki University of Technology,
Finland.
\end{acknowledgments}

\appendix
\section{Vortex Formation and Berry Phase}
It is shown that the formation of a vortex in
our scenario is understood
from a slightly different viewpoint. Namely, we show that the phase appearing
in the end of the process may be identified with the Berry phase
associated with the adiabatic change of the magnetic field.

Let a point on the unit sphere in Fig. 11 (a) denote the hyperfine
spin state with $F=1$. All the spins near the origin $r=0$ have
$F_z \simeq -1$ at $t=0$
and hence correspond to the points near the south pole. Now the axial
field $B_z$ monotonically decreases such that eventually all the
spins near the origin have $F_z \simeq +1$ at $t=T$ and hence are expressed by
the points near the north pole. The path a spin follows on the unit
sphere depends on the position of the spin relative to the origin.
Figure 11 (b) shows the configuration of four
hyperfine spins $\bv{F}$ at $t=T/2$, when
$B_z$ vanishes and $\bv{F}$ is determined by the quadrupole field.
The path $C_1$ in Fig. 11 (a) shows the trajectory the spin 1 in Fig. 11 (b)
follows when $t$ is varied from $0\ (\beta =0)$ to $T\ (\beta=\pi)$. Similarly
the path $C_2$ in Fig. 11 (a) is the trajectory of the spin 2 in Fig. 11 (b),
and so on. When the spins 1 and 2
left the south pole at $t=0$, they had the common phase
factor, while at $t=T$ they obtain the relative phase equal to the 
solid angle subtended by the trajectories $C_1$ and $C_2$.
The shaded area in Fig. 11 (a) shows this area which subtends
the solid angle $\pi$. Similarly the paths 1 and 3 (1 and 4) subtend the
solid angle $2 \pi$ ($3 \pi$), resulting in the relative phase $2\pi
(3\pi)$ between the spins 1 and 3 (1 and 4), respectively,
when $B_z$ is completely reversed. Accordingly, as one completes
a loop surrounding the origin, one measures the phase change of $4\pi$
observing that a vortex formation has taken place. 
\begin{figure}
\begin{center}
\includegraphics[width=8cm]{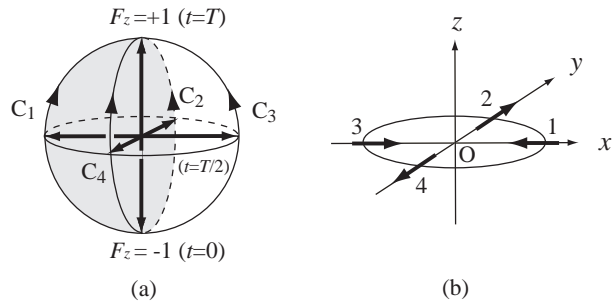}
\end{center}
\caption{Formation of a vortex in the present scenario may be
understood in terms of the Berry phase associated with each
spin. When $|B_z| \gg |B_{\perp}|$ at $t=0$, all the spins near the
origin have $F_z \simeq -1$ with the same phase. Due to the ``twisting''
of the spins during the evolution, they obtain different phases
depending on the trajectories they follow. (b) shows four spins
at $t=T/2$ where $B_z$ vanishes. The spin 1 follows the trajectory
denoted by $C_1$ in (a) starting from the south pole and ending
at the north pole. Similarly the spins 2, 3 and 4 in (b) follows
the trajectories $C_2, C_3$ and $C_4$ in (a), respectively.
The shaded area $(\pi/2 \leq \phi\leq \pi, 0 \leq \beta \leq \pi)$
in (a) is the solid angle subtended by the paths $C_1$ and $C_2$.
Since this area measures $\pi$, the spin $2$ has a phase
$\pi$ relative to that of the spin $1$ at $t=T$.
Similarly the spins 3 and 4 obtain phases $2\pi$ and $3\pi$, respectively,
relative to that of the spin 1 at $t=T$.}
\label{fig:9}
\end{figure}


\end{document}